\documentclass[showpacs,preprintnumbers,twocolumn,prd]{revtex4}
\usepackage{amsmath,amssymb}
\usepackage[dvips]{graphicx}

\newcommand{\ubu}{\langle\bar{u}u\rangle}
\newcommand{\dbd}{\langle\bar{d}d\rangle}
\newcommand{\sbs}{\langle\bar{s}s\rangle}
\newcommand{\rmd}{\mathrm{d}}
\newcommand{\rme}{\mathrm{e}}
\newcommand{\rmi}{\mathrm{i}}
\newcommand{\tr}{\mathrm{tr}}
\newcommand{\gs}{g_{\text{S}}}
\newcommand{\gv}{g_{\text{V}}}
\newcommand{\gd}{g_{\text{D}}}
\newcommand{\MeV}{\;\text{MeV}}
\newcommand{\Op}{\Omega_{\text{PNJL}}}
\newcommand{\lb}{\bar{\ell}}
\newcommand{\Lag}{\mathcal{L}}
\newcommand{\UA}{\mathrm{U_A}(1)}

\begin{document}
\title{Phase diagrams in the three-flavor Nambu--Jona-Lasinio model
  with the Polyakov loop}
\author{Kenji Fukushima}
\affiliation{Yukawa Institute for Theoretical Physics,
 Kyoto University, Kyoto 606-8502, Japan}
\begin{abstract}
 We present extensive studies on hot and dense quark matter with two
 light and one heavy flavors in the Nambu--Jona-Lasinio model with the
 Polyakov loop (so-called PNJL model).  First we discuss prescription
 dependence in choosing the Polyakov loop effective potential and
 propose a simple and rather sensible ansatz.  We look over
 quantitative comparison to the lattice measurement to confirm that
 the model captures thermodynamic properties correctly.  We then
 analyze the phase structure by changing the temperature, quark
 chemical potential, quark masses, and coupling constants.  We
 particularly investigate how the effective $\UA$ restoration and the
 induced vector-channel interaction at finite density would affect the
 QCD critical point.
\end{abstract}
\preprint{YITP-08-19}
\pacs{12.38.Aw, 11.10.Wx, 11.30.Rd, 12.38.Gc}
\maketitle


\section{INTRODUCTION}

  The phase diagram of hot and dense matter out of quarks and gluons
described by Quantum Chromodynamics (QCD) has attracted theoretical
and experimental interest for decades~\cite{early,review}.  We can
define one phase transition associated with chiral symmetry
restoration in the vanishing quark mass limit (i.e.\ $m_q\to0$) which
is commonly referred to as the chiral phase transition.  In the
quenched limit with infinitely heavy quark mass (i.e.\
$m_q\to\infty$), on the other hand, we can define another phase
transition from the hadron (glueball) phase to the color deconfinement
phase.  The question is then the nature of these phase transitions
with intermediate quark masses of two light (\textit{up} and
\textit{down}) and one heavy (\textit{strange}) flavors.  We stress
that the chiral and deconfinement phase transitions are conceptually
distinct phenomena and, theoretically speaking, they reside in the
opposite limits with respect to the quark mass.  Nevertheless, the
standard QCD phase diagram on the plane of the temperature $T$ and the
chemical potential $\mu$ has only a single transition or crossover
boundary.  Whether this is really the case is not trivial
\textit{a priori} and not quite settled yet.

  It is the result from the Monte-Carlo integration of the (quenched)
QCD partition function on the lattice that had led us to this phase
diagram with a single phase boundary~\cite{Kogut:1982rt}.  (See
Ref.~\cite{Baym:2001in} and references therein for historical
background.)  Later on, the lattice QCD simulation with dynamical
quarks~\cite{Fukugita:1986rr} have confirmed that the chiral and
deconfinement phase transitions occur at the same temperature (or at
different but close temperatures~\cite{Aoki:2006br}).  This
observation suggests that two phenomena of chiral restoration and
color deconfinement should be locked by some dynamical mechanism so
that they should take place (nearly) at once.

  We can find the first successful study based on dynamical model to
give an account for this locking mechanism in the work by Gocksch and
Ogilvie~\cite{Gocksch:1984yk}.  They have constructed the effective
action of QCD by means of the strong coupling and large dimensional
expansions.  The same action has been discussed at finite $T$ and
$\mu$ also by Ilgenfritz and Kripfgantz~\cite{Ilgenfritz:1984ff}.
There were proposed some generic mixing arguments which aim to explain
the coincidence of critical temperatures in a model independent
way~\cite{mixing}.  The mixing effect, however, does not suffice to
force two crossovers to be a single one in view of the associated
peaks in the susceptibility;  two separate crossovers (susceptibility
peaks) with mixing for each cannot be ruled out.  That means, the
mixing effect is a necessary but not sufficient condition in order to
realize the coincidence in a way seen on the lattice~\cite{hatta}.
Thus, the locking between chiral restoration and deconfinement should
need more tangled dynamical properties of two phenomena.

  To reveal the relevant dynamics, the present author proposed a
useful model~\cite{Fukushima:2003fw} based on the Nambu--Jona-Lasinio
(NJL) model~\cite{Klevansky:1992qe,Hatsuda:1994pi} with the Polyakov
loop degrees of freedom augmented, which was inspired by the strong
coupling
analyses~\cite{Gocksch:1984yk,Ilgenfritz:1984ff,Fukushima:2002ew}.
The peculiar feature in this model is that we can uniquely determine
the coupling between the chiral condensate, which is an order
parameter for the chiral phase transition in $m_q\to0$, and the
Polyakov loop, which is an order parameter for the deconfinement phase
transition in $m_q\to\infty$.  The model inputs and outputs have been
carefully compared to the lattice QCD data by Ratti, Thaler, and
Weise~\cite{Ratti:2005jh} and they named this hybrid description as
the \textit{PNJL model}.

  The PNJL model has been generalized to the three-flavor case
recently~\cite{Fu:2007xc,Ciminale:2007sr}.  In the present paper we
shall extensively explore the phase diagrams in the three-flavor PNJL
model by changing four physical variables, namely, $T$, $\mu$, the
light quark mass $m_{ud}$, and the heavy quark mass $m_s$.  First we
revisit the choice of the Polyakov loop effective potential that
cannot avoid ambiguity in the PNJL model approach.  We claim that a
careful consideration is necessary for the effective potential form.
Once we fix the pure gluonic sector by specifying the potential, we
can calculate the mean-fields of the chiral condensate and the
Polyakov loop to draw the phase diagrams.

  Although it is usually assumed implicitly, we have no
first-principle insight into the locking of two crossovers in the
finite-density region.  The lattice Monte-Carlo simulation is of no
practical use except when $\mu$ is much smaller than $T$.  So far it
seems that almost nothing but the PNJL model can access both
transitions at any density.  Strictly speaking, in fact, the
mean-field treatment of the PNJL model is not totally free from the
sign problem.  Detailed analyses in Ref.~\cite{Fukushima:2006uv}
support that the saddle-point of the mean-field energy leads to an
appropriate estimate for the mean-fields, however.  Hence, we will not
argue the sign problem any more in the present work.

  In this paper, after we check that our results from the three-flavor
PNJL model are consistent with the state-of-art lattice simulation at
zero chemical potential~\cite{Cheng:2007jq}, we will shift our
emphasis toward the QCD (chiral) critical end-point in the last half.
It should be noted that the terminal point of the first-order phase
boundary has a second-order phase transition characterized by the
universality class of the Z(2) Ising model and this special point is
often called the QCD critical point.  The search for the critical
point is one of the most interesting problems in finite density QCD
because it provides us with a firm milestone for our quest for the QCD
phase diagram.  If we are lucky enough to find out the critical point
as predicted in theory, we can get confident about our theory
reliability.  This is, so to speak, a mutual correspondence between
theory and experiment, which is an ideal situation for sound
scientific developments.

  The existence of the QCD critical point has not been established
yet.  We cannot exclude a possibility that the QCD phase transition is
smooth everywhere in the $\mu$-$T$ plane, while there are a pile of
indirect evidences for its
existence~\cite{Asakawa:1989bq,Barducci:1989wi,Berges:1998rc,%
Halasz:1998qr,Hatta:2002sj,Fodor:2001pe,Gavai:2004sd}.  In model
studies, in fact, a minor modification in the treatment could easily
smear a first-order phase transition out into a crossover, as
demonstrated later.  In particular we shall pay attention to two
obscure factors which may significantly affect where the critical
point is and even whether it exists.  Those two factors are the
magnitude of the $\UA$-breaking anomaly interaction and the
vector-channel interaction.  The former, the $\UA$-breaking term,
induces a six-quark vertex called the 't~Hooft term which mixes three
different flavors up and is responsible for the first-order phase
transition in the chiral limit~\cite{Pisarski:1983ms}.  It could be
possible at finite temperature and density that the 't~Hooft
interaction is reduced by instanton
suppression~\cite{Shuryak:1993ee,Cohen:1996ng,Fukushima:2001hr}.  The
latter effect, i.e.\ the vector-channel interaction term, does not
break chiral symmetry and the zeroth component directly couples the
quark density.  It is thus likely that the finite-density environment
enhances or induces interactions in the vector channel which could
weaken the first-order phase
transition~\cite{Asakawa:1989bq,Klimt:1990ws,Kitazawa:2002bc,Sasaki:2006ws}.
In this paper we shall quantify these effects on the location of the
QCD critical point using the three-flavor PNJL model.


\section{MODEL SETUP}

  The present author proposed the PNJL model action in
Ref.~\cite{Fukushima:2003fw} inspired by the effective action in
strong-coupling QCD with dynamical
quarks~\cite{Gocksch:1984yk,Ilgenfritz:1984ff,Fukushima:2002ew}.  It
is possible to some extent to elaborate a field-theoretical setup for
the PNJL model starting with the Lagrangian
density~\cite{Ratti:2005jh}.  For this purpose it is required to
assume a homogeneous mean-field distribution of the Polyakov loop.
In other words, the temporal component of the gauge field, $A_4$, in
Euclidean space-time must be approximated by a spatially constant
mean-field, so that one can perform the one-loop integration with
respect to thermal quarks in a closed form.  This thermal integration
leads to the unique coupling between the chiral condensate and the
Polyakov loop.  Spatial uniformity is in fact a mean-field ansatz,
however, and it makes a contrast to the strong-coupling
framework~\cite{Fukushima:2002ew}, as we shall discuss shortly.

  In the PNJL model the Polyakov loop is therefore put in as a global
mean-field rather than a local dynamical variable, which is analogous
to the treatment of the chiral condensate in the ordinary NJL model;
the Lagrangian density with a shift by the mean-field is sometimes
referred to as the mean-field Lagrangian that contains no kinetic
term for the mean-field.  Such an approximation should work to
investigate the bulk property of the thermodynamic system, while we
have to be aware that the mean-field model cannot properly deal with
the spatial structure of confined objects.  It is beyond the scope of
the simple PNJL model framework, for instance, to extract the
heavy-quark potential.

  All the model ingredients are thus given as mean-field variables.
Here we would prefer to start with the mean-field free-energy after
one-loop integration for the model setup.  Let us decompose the
free-energy below into four pieces and discuss them in order.  That
is, the total free-energy (or the grand potential) is a sum of four
contributions;
\begin{equation}
 \Op = \underbrace{\Omega_{\text{cond}} + \Omega_{\text{zero}}
  + \Omega_{\text{quark}}}_{\text{NJL part}} \;\;+\!\!
  \underbrace{\Omega_{\text{Polyakov}}}_{\text{pure gluonic part}}
  \!\!\! \,,
\label{eq:total_energy}
\end{equation}
where $\Omega_{\text{cond}}$ represents the condensation energy in the
chiral sector, $\Omega_{\text{zero}}$ the zero-point energy which is
important in the NJL model formulation, $\Omega_{\text{quark}}$ the
thermal quark contribution with the Polyakov loop coupling coming from
the Dirac determinant, and $\Omega_{\text{Polyakov}}$ gives the
effective potential in terms of the Polyakov loop variable.  As
indicated in Eq.~(\ref{eq:total_energy}), we can deduce the first
three from the standard NJL model and the last one from the pure
gluonic theory.


\subsection{Condensation Energy}

  We can read the condensation energy from the standard NJL model
Lagrangian.  Using the notation by Hatsuda and
Kunihiro~\cite{Hatsuda:1994pi}, we write the four-quark interaction
in the scalar channel and the six-quark 't~Hooft interaction as
\begin{equation}
 \Lag_S = \frac{\gs}{2}\bigl[(\bar{\psi}\lambda_a\psi)^2
  +(\bar{\psi}\rmi\gamma_5\lambda_a\psi)^2 \bigr] \,,
\end{equation}
and
\begin{equation}
 \Lag_A = \gd\bigl[\det\bar{\psi}(1-\gamma_5)\psi +
  \text{h.c.}\bigr]\,,
\end{equation}
respectively.  For later convenience we also give an expression for
the vector-channel interaction;
\begin{equation}
 \Lag_V = -\gv\,(\bar{\psi}\gamma_\mu\psi)^2 \,.
\label{eq:vec}
\end{equation}
For the moment we will work only in the $\gv=0$ case.  Here,
$\lambda_a$'s are the Gell-Mann matrices in flavor space (with
$\lambda_0=\sqrt{2/3}$) and the matrix determinant is taken also in
flavor space.  In the mean-field approximation with three condensates,
$\ubu$, $\dbd$, and $\sbs$, the scalar four-quark interaction is
rewritten as
\begin{align}
 \gs(\bar{u}u)^2 &\to \gs\bigl(\bar{u}u-\ubu+\ubu\bigr)
  \bigl(\bar{u}u-\ubu+\ubu\bigr) \notag\\
 &\simeq \gs\ubu^2 +2\gs\ubu\bigl(\bar{u}u-\ubu\bigr) \notag\\
 &= -\gs\ubu^2 + 2\gs\ubu\bar{u}u \,,
\label{eq:ex_u}
\end{align}
in the $u$-quark sector and likewise for $d$-quarks and $s$-quarks.
In this way we can readily reach the following expression for the
condensation energy;
\begin{equation}
 \Omega_{\text{cond}} = \gs\bigl(\ubu^2\!+\dbd^2\!+\sbs^2\bigr)
  + 4\gd\ubu\dbd\sbs \,.
\end{equation}
We see that the six-quark interaction induces the flavor-mixing
interaction indeed which makes the phase transition of first-order in
the presence of massless three flavors.


\subsection{Zero-Point Energy}

  The zero-point energy diverges and requires the ultraviolet cutoff
$\Lambda$ to regularize the three-momentum integration.  Since the NJL
model is a non-renormalizable cutoff theory depending on the choice of
$\Lambda$, the zero-point energy contribution largely affects the
model output.  With the quasi-quark energy dispersion relation,
$\varepsilon_i(p)=\sqrt{p^2+M_i^2}$, the zero-point energy can be
expressed simply as a summation of all $\varepsilon_i(p)/2$, that is,
\begin{equation}
 \Omega_{\text{zero}} = -2N_c\sum_i\int^\Lambda\!
  \frac{\rmd^3p}{(2\pi)^3}\;\varepsilon_i(p) \,,
\end{equation}
where $2$ is the spin degrees of freedom, $N_c=3$ is the number of
colors, and the particle and anti-particle contributions cancel $2$ in
the denominator of $\varepsilon_i(p)/2$.  The constituent quark mass
is defined as a sum of the current quark mass and the mean-field as
\begin{equation}
 \begin{split}
 M_u &= m_u -2\gs \ubu -2\gd \dbd\sbs \,,\\
 M_d &= m_d -2\gs \dbd -2\gd \sbs\ubu \,,\\
 M_s &= m_s -2\gs \sbs -2\gd \ubu\dbd \,,
 \end{split}
\end{equation}
which is understood from the second term in Eq.~(\ref{eq:ex_u}).


\subsection{Thermal Quark Energy}

  The thermal quark energy is where we can uniquely introduce coupling
between the chiral condensate and the Polyakov loop.  In the PNJL
model, under the assumption of the presence of the spatially uniform
Polyakov loop background, the one-loop free-energy is modified as
\begin{align}
 \Omega_{\text{quark}} =& -2T\sum_i\int\!
  \frac{\rmd^3p}{(2\pi)^3}\,\Bigl\{\ln\det\Bigl[1+L\,
  \rme^{-(\varepsilon_i(p)-\mu)/T}\Bigr]\notag\\
 &\quad +\ln\det\Bigl[1+L^\dagger\rme^{-(\varepsilon_i(p)+\mu)/T}\Bigr]
  \Bigr\} \,.
\end{align}
Let us comment on preceding
works~\cite{Chandrasekharan:1995nf,Meisinger:1995ih} in which a
similar coupling form is addressed.

  We note that the above expression is identical with that in the
strong coupling expansion but the physics content is slightly
different.  The Polyakov loop $L$ is a mean-field from the beginning
here, whereas the strong coupling calculation at finite temperature
decouples the temporal hopping from spatial link
variables~\cite{Fukushima:2002ew}.  As a result, the quark excitation
is static in the strong-coupling leading order, and the above
expression results at each lattice site in this way, that is, $L$
could be a local variable in the strong coupling expansion.

  It is noteworthy that the three-momentum integration above is finite
and has no need for the ultraviolet cutoff.  We can thus relax the
cutoff in the thermal quark energy, though we found that the $s$-quark
sector behaves unnaturally at extremely high temperature without the
cutoff, which is of no importance practically.  In this work we will
not impose the momentum cutoff onto the thermal quark energy in order
to let the thermodynamic quantities free from cutoff artifact.

  The Polyakov loop $L$ is an $N_c\times N_c$ matrix in color space
and is defined originally in terms of $A_4$.  The explicit form of the
Polyakov loop is irrelevant in our study because we treat it as a
model variable and will not return to the original definition of the
Polyakov loop in terms of the gauge field.

  In the simplest mean-field approximation one can express the
free-energy as a function of the traced Polyakov loop expectation
value defined by
\begin{equation}
 \ell = \frac{1}{N_c}\langle\tr L\rangle \,,\quad
 \lb = \frac{1}{N_c}\langle\tr L^\dagger\rangle \,.
\end{equation}
It should be mentioned that we must distinguish $\ell$ and $\lb$ at
finite density~\cite{Fukushima:2006uv,Allton:2002zi,Dumitru:2005ng};
both $\ell$ and $\lb$ are real, and nevertheless, $\lb>\ell$ whenever
$\mu>0$.  This is because a finite chemical potential gives rise to a
$C$-odd term like $\mu\,\mathrm{Im}[\tr L]$ in the average weight
leading to $\lb-\ell\sim \mu\,\langle(\mathrm{Im}[\tr L])^2\rangle>0$
for small $\mu$.  We can also give an intuitive explanation;  $\lb$
represents the exponential of the free-energy gain, $f_{\lb}$, by the
presence of an anti-quark.  The test charge brought in by an
anti-quark can be more easily screened in a medium with more quarks
than anti-quarks.  Therefore, $f_{\lb} < f_{\ell}$, that means,
$\lb=\rme^{-f_{\lb}/T} > \ell=\rme^{-f_{\ell}/T}$ for a positive
$\mu$.

  It is straightforward to take an average of the $3\times3$
determinant to reach
\begin{align}
 & \Bigl\langle \det\bigl[1+L\,\rme^{-(\varepsilon-\mu)/T}\bigr]
  \Bigr\rangle = 1 + \rme^{-3(\varepsilon-\mu)/T} \notag\\
 &\qquad\qquad\qquad +3\,\ell\,\rme^{-(\varepsilon-\mu)/T}
   +3\,\lb\,\rme^{-2(\varepsilon-\mu)/T} \,,
\label{eq:det_lp} \\
 & \Bigl\langle \det\bigl[1+L^\dagger\,\rme^{-(\varepsilon+\mu)/T}\bigr]
  \Bigr\rangle = 1 + \rme^{-3(\varepsilon+\mu)/T} \notag\\
 &\qquad\qquad\qquad +3\,\lb\,\rme^{-(\varepsilon+\mu)/T}
   +3\,\ell\,\rme^{-2(\varepsilon+\mu)/T} \,.
\label{eq:det_la}
\end{align}
In this work we use the logarithm of the above expressions as the
mean-field free-energy and will not perform the group integration over
$L$.  Roughly speaking, the approximation involving the group
integration~\cite{Gocksch:1984yk} corresponds to what is called the
Weiss mean-field approximation in the spin system.  The integration
has an effect on the quantitative results~\cite{Megias:2004hj} but a
simple mean-field treatment suffices for our present purpose.  We also
remark that the action is invariant under simultaneous replacement
$\ell\leftrightarrow\lb$ and $-\mu\leftrightarrow+\mu$.


\subsection{Polyakov Loop Energy}

  In the definition of the PNJL model the choice of the Polyakov loop
potential has subtlety because the effective potential has not been
known directly from the lattice QCD simulation.  In the present study
we will assume the strong-coupling inspired form of
\begin{equation}
 \begin{split}
 & \Omega_{\text{Polyakov}} = -b\cdot T\, \Bigl\{54\,
  \rme^{-a/T}\,\ell\,\lb \\
 &\qquad +\ln\bigl[1-6\,\ell\,\lb -3(\,\ell\,\lb\,)^2
  +4(\,\ell^3+\lb\,^3)\bigr] \Bigr\} \,.
 \end{split}
\label{eq:pot_gl}
\end{equation}
The logarithmic term appears from the Haar measure of the group
integration with respect to the SU(3) Polyakov loop matrix.  The first
term is reminiscent of the nearest neighbor interaction in the
effective action at strong coupling.  The temperature-dependent
coefficient of this $\ell\bar{\ell}$ term controls the deconfinement
phase transition temperature.  It should be, however, noted that the
model parameters are assumed to be temperature-independent.  (See
Ref.~\cite{Schaefer:2007pw} for the running coupling effects including
renormalization.)

  In this simple ansatz for the Polyakov loop potential, we have two
parameters; $a$ and $b$.  The deconfinement phase transition is
determined solely by the choice of $a$, while $b$ parametrizes the
relative strength of mixing between the chiral and deconfinement phase
transitions.  If $b$ is small, chiral restoration dominates the phase
transition, and if $b$ is large, deconfinement is more governing.

  We will numerically make a comparison between the above-proposed
ansatz and others in the next section.


\section{NUMERICAL PROCEDURES}

  Now that we have specified all the constituents in the model action,
we get ready to proceed to the numerical analyses.  We will solve the
following four equations in a self-consistent way,
\begin{equation}
 \frac{\partial\Op}{\partial\ubu} = \frac{\partial\Op}{\partial\sbs}
  =\frac{\partial\Op}{\partial\ell}=\frac{\partial\Op}{\partial \lb}
  =0
\label{eq:gap_eq}
\end{equation}
to acquire $\ubu=\dbd$, $\sbs$, $\ell$, and $\lb$ as functions of the
model input.  For this purpose we have to fix all the model
parameters, $\Lambda$, $\gs$, $\gd$ in the NJL potential, and $a$ and
$b$ in the Polyakov loop potential.


\subsection{Parameter Choice}

  The Polyakov loop coupling appears only in the thermal part, that
means that the NJL model parameters fixed at $T=\mu=0$ are not
affected by introduction of the Polyakov loop coupling. In this work
we will employ the widely accepted parameter set according to Hatsuda
and Kunihiro~\cite{Hatsuda:1994pi};
\begin{equation}
 \begin{split}
 & \Lambda = 631.4\MeV\,,\\
 & m_{ud}=5.5\MeV\,,\quad m_s = 135.7\MeV\,, \\
 & \gs\cdot\Lambda^2 = 3.67\,,\qquad \gd\cdot\Lambda^5 = -9.29 \,,
 \end{split}
\end{equation}
which nicely reproduces the $\pi$ mass, the $K$ mass, the $\eta'$
mass, and the $\pi$ decay constant $f_\pi$.  Here $m_{ud}$ stands
representatively for the light quark mass, i.e.\ $m_{ud}=m_u=m_d$.

  Regarding the Polyakov loop potential, we can fix the parameter $a$
by the condition that the first-order phase transition in the pure
gluodynamics takes place at $T=270\MeV$, which yields
\begin{equation}
 a = 664\MeV \,,
\end{equation}
and then the remaining variable is $b$ only.  Actually, the
determination of $b$ suffers uncertainty and there is no established
prescription.  In this study we shall take a value of $b$ that leads
to simultaneous crossovers of chiral restoration and deconfinement
around $T\simeq 200\MeV$. As a result, we set
\begin{equation}
 b\cdot\Lambda^{-3} = 0.03 \,.
\label{eq:b}
\end{equation}


\subsection{Other Polyakov Loop Potentials}

  The choice of the Polyakov loop potential has some variations, as we
have mentioned before.  Our choice of Eq.~(\ref{eq:pot_gl}) is much
simpler than the widely accepted forms by Ratti, Thaler, and
Weise~\cite{Ratti:2005jh} and by R\"{o}{\ss}ner, Ratti, and
Weise~\cite{Roessner:2006xn}.  It would be instructive to scrutinize
respective forms and quantify the difference numerically.  Let us call
the ``RTW05 potential'' to indicate
\begin{equation}
 \Omega_{\text{RTW05}} = T^4 \biggl[ -\frac{b_2(T)}{2}\ell\lb
  -\frac{b_3}{6}\bigl(\ell^3\!+\!\lb^3\bigr) +\frac{b_4}{4}
  \bigl(\ell\lb\bigr)^2 \biggr]
\label{eq:RTW05}
\end{equation}
with $b_2(T)=a_0+a_1(T_0/T)+a_2(T_0/T)^2+a_3(T_0/T)^3$, which is
proposed in Ref.~\cite{Ratti:2005jh}.  There are seven parameters,
$a_0=6.76$, $a_1=-1.95$, $a_2=2.625$, $a_3=-7.44$, $b_3=0.75$,
$b_4=7.5$, and $T_0=270\MeV$ such that the potential~(\ref{eq:RTW05})
reproduces the pressure, energy density, and entropy density in the
pure gluonic sector measured on the lattice.  A slightly different
choice is suggested in Ref.~\cite{Roessner:2006xn} which we shall
call the ``RRW06 potential'';
\begin{equation}
 \begin{split}
 & \Omega_{\text{RRW06}} = T^4\biggl\{ -\frac{a(T)}{2}\ell\lb \\
 &\qquad +b(T)\ln\bigl[ 1-6\ell\lb-3(\ell\lb)^2+4(\ell^3+\lb^3)\bigr]
  \biggr\}
 \end{split}
\label{eq:RRW06}
\end{equation}
with $a(T)=a_0+a_1(T_0/T)+a_2(T_0/T)^2$ and $b(T)=b_3(T_0/T)^3$.  Five
parameters are fixed as $a_0=3.51$, $a_1=-2.47$, $a_2=15.2$,
$b_3=-1.75$, and $T_0=270\MeV$.  We note that $b_3$ plays the same
role as $b$ in our ansatz~(\ref{eq:pot_gl}).  Actually, if we
substitute $T_0=190\MeV$ to lower the crossover temperature as argued
in Ref.~\cite{Ratti:2005jh}, $b_3T_0^3\cdot\Lambda^3\simeq 0.044$
(where $\Lambda$ is not our value but $650\MeV$ used in
Ref.~\cite{Ratti:2005jh}) which turns out to be comparable to our
choice~(\ref{eq:b}).

  Under the assumption that $\Omega_{\text{RTW05}}$ and
$\Omega_{\text{RRW06}}$ correspond to the \textit{total} negative
pressure in the pure gluonic theory, they approach the
Stefan-Boltzmann limit at high temperature, that is,
$p=(2\cdot8\cdot\pi^2/90)T^4=1.75T^4$.  One can easily make this sure
from $-a_0/2-b_3/3+b_4/4=-1.75$ in $\Omega_{\text{RTW05}}$ and
$-a_0/2=-1.75$ in $\Omega_{\text{RRW06}}$.

  We would claim, however, that $\Omega_{\text{RTW05}}$ and
$\Omega_{\text{RRW06}}$ might overcount the relevant degrees of
freedom in the system.  In the high temperature limit not only the
Polyakov loop but also the deconfined transverse gluons contribute to
the pressure.  Since the Polyakov loop corresponds to the longitudinal
gauge field, the Stefan-Boltzmann limit should be saturated by the
transverse gluons but not the Polyakov loop.  It is thus a subtle
assumption that the effective potential with respect to the order
parameter field can reproduce the total pressure, energy density, and
entropy density for all temperatures.

  One can understand this from a more familiar example.  Let us
consider the mean-field effective potential in the O(4) linear sigma
model.  The effective potential with respect to the $\sigma$
condensate describes the chiral phase transition.  The total pressure
should contain contributions from the $\pi$ excitations too which are
not fully included in the effective potential in terms of
$\langle\sigma\rangle$.

  It is not our point to insist that $\Omega_{\text{RTW05}}$ and
$\Omega_{\text{RRW06}}$ are doubtful.  Our main point lies in the
other way around in fact.  We presume that their parametrization works
in effect for the following reason;  the pressure contribution from
transverse gluons is a function of $T$, and the Polyakov loop is also
a function of $T$, and so the former can be implicitly parametrized by
the latter.  Then, it is possible to express the total pressure in the
form of Eq.~(\ref{eq:RTW05}) or (\ref{eq:RRW06}).  One has to keep in
mind, however, that the total pressure in this interpretation would
make sense provided that the Polyakov loop is already solved as a
function of $T$.  Therefore, one should solve Eq.~(\ref{eq:gap_eq})
first and then one can fit the total pressure using
Eq.~(\ref{eq:RTW05}) or (\ref{eq:RRW06}) with solved $\ell(T)$ and
$\lb(T)$ substituted.  One should not use the total pressure itself to
optimize the variational parameters $\ell$ and $\lb$.  This may
explain why the critical temperature determined with
Eq.~(\ref{eq:RTW05}) or (\ref{eq:RRW06}) put into the gap equations
becomes relatively higher.  The Polyakov loop effective potential
which overcounts the gluonic degrees of freedom would drag the
crossover point closer to the pure gluonic transition temperature
$T_0=270\MeV$.
 
\begin{figure}
 \includegraphics[width=7.5cm]{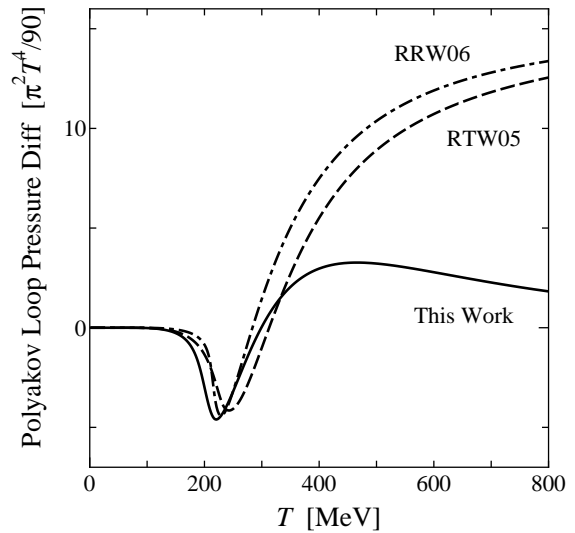}
 \caption{Comparison of the Polyakov loop pressure excess as a function
 of the temperature.  The vertical axis signifies the effective
 degrees of freedom.  RTW05 and RRW06 represent
 $\Omega_{\text{RTW05}}$ and $\Omega_{\text{RRW06}}$ with $\ell$ and
 $\lb$ given as a solution of the \textit{full} gap equation with two
 quark flavors.  For comparison we used the same NJL model parameters,
 $\Lambda$, $\gs$, and $m_u=m_d$ as in Ref.~\cite{Ratti:2005jh} and
 quenched the $s$-quark sector to draw the solid curve which
 represents our $\Omega_{\text{Polyakov}}$.}
 \label{fig:gl}
\end{figure}

  We emphasize that our simple choice of the Polyakov loop potential
is physically natural and, interestingly, it makes only little
difference from the numerical results based on the RTW05 or RRW06
potential.  This sounds very good, for our new potential ansatz does
not ruin the nice agreement to the lattice data addressed in
Refs.~\cite{Ratti:2005jh,Roessner:2006xn,Ratti:2007jf}.  In
Fig.~\ref{fig:gl} we plot the Polyakov loop pressure difference from
the zero temperature value using the mean-fields obtained from the
full gap equations with two flavors.  We could, of course, show the
genuine total pressure with the Polyakov loop and quark contributions
both.  We subtracted the quark contribution in Fig.~\ref{fig:gl}
because the quark contribution makes the comparison blurred;  the
quasi-quark pressure is dominating up to near $T_c$ but it is not
sensitive to the choice of the Polyakov loop with
$\Omega_{\text{quark}}$ untouched.  The non-trivial part is the
Polyakov loop contribution that we now focus on here.

  In the absence of interaction, the pressure is given by the
Stefan-Boltzmann law, $\pi^2 T^4/90$, multiplied by the effective
degrees of freedom which we denote by $\nu$.  To see how $\nu$
increases as $T$ goes up, we normalize the pressure by the
Stefan-Boltzmann unit; $\pi^2 T^4/90$.  Clearly both
$\Omega_{\text{RTW05}}$ and $\Omega_{\text{RRW06}}$ increase with
increasing $T$ and asymptotically approach the value of $\nu=$ 2
(polarization) $\times$ 8 (color)=16.  It is so by construction, as we
explained.  It is intriguing to note that our ansatz~(\ref{eq:pot_gl})
results in the solid curve in Fig.~\ref{fig:gl} which is close to the
dashed and dot-dashed curves by $\Omega_{\text{RTW05}}$ and
$\Omega_{\text{RRW06}}$ as long as the temperature is below
$300\MeV\simeq 1.5T_c$.  We do not have to care much about the
discrepancy in the higher temperature region, in fact, because the
validity region in the present study extends at best up to $\sim 2T_c$
above which transverse gluons should be dominant.  Therefore, we can
conclude that all these potential choices are consistent to each other
within the validity range of the temperature.  In our
choice~(\ref{eq:pot_gl}) the effective degrees of freedom slowly
decrease at higher temperature in the Stefan-Boltzmann unit.  This is
reasonable because the Polyakov loop must give way to transverse
gluons.

  The nearly coincidence of three curves in the vicinity of $T_c$ in
Fig.~\ref{fig:gl} delivers us an important message.  The Polyakov loop
takes on a major fraction of the system pressure up to the temperature
around $1.5T_c$.  We should recall that two parameters, $a$ and $b$,
in Eq.~(\ref{eq:pot_gl}) have been fixed not to reproduce the pressure
but just to yield $T_0=270\MeV$ in the pure gluonic sector and
$T_c\simeq200\MeV$ with $2+1$ flavors.


\section{Zero Density Results}

  Here we show the model results at zero quark density with our
choice of the model parameters.  In our subsequent discussions we will
make clear the virtues of the PNJL model as well as some caveats.


\subsection{Order Parameters}

  Because nothing breaks isospin symmetry in this work, we will show
the numerical results only for the $u$-quark sector which is
degenerate to the $d$-quark sector.

\begin{figure}
 \includegraphics[width=7.5cm]{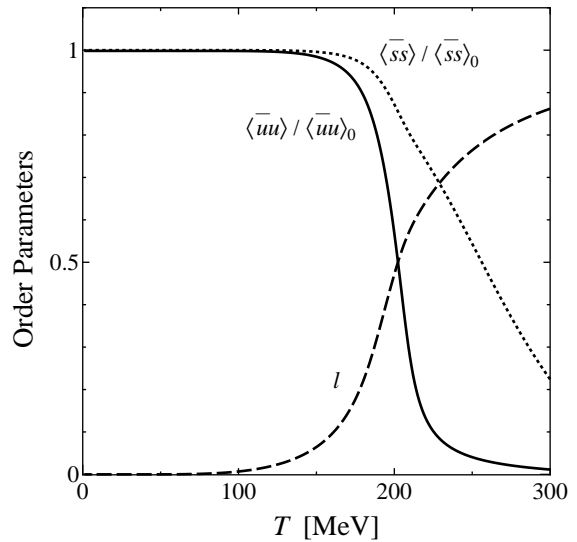}
 \caption{Order parameters at zero density as a function of the
 temperature.  The chiral condensates are normalized by
 $\ubu_0=(246\MeV)^3$ and $\sbs_0=(267\MeV)^3$.  The solid and dotted
 curves represent the $u$-quark and $s$-quark chiral condensates,
 respectively, and the dashed curve represents the traced Polyakov
 loop $\ell$.}
 \label{fig:chiral_plane}
\end{figure}

  First of all, we present Fig.~\ref{fig:chiral_plane} to confirm that
simultaneous crossovers of deconfinement and chiral restoration
certainly realize in the PNJL model.  The chiral condensates are
normalized by their zero-temperature value; $\ubu_0=(246\MeV)^3$ and
$\sbs_0=(267\MeV)^3$ for light and heavy quarks, respectively.

  The reason why we find the simultaneous crossovers around
$T_c\simeq 200\MeV$ (the temperature derivative gives
$T_c=204.8\MeV$) is that we have chosen the value of $b$ as
Eq.~(\ref{eq:b}) to adjust the crossover temperature by hand.  Thus,
we note that the crossover temperature is not the model output but the
input.  Nevertheless we would comment on a non-trivial feature
inherent in the model dynamics;  the chiral phase transition can never
occur until the Polyakov loop grows up~\cite{Fukushima:2003fw}.  It is
also interesting to look at the behavior of the $s$-quark chiral
condensate depicted by the dotted curve.  The results for $\sbs$ are
the output rather than the input unlike $\ubu$.  If we define a
crossover temperature for the $s$-quark sector, it should be higher
than the simultaneous crossovers due to the explicit breaking of
chiral symmetry.


\subsection{Effective Confinement}

\begin{figure}
 \includegraphics[width=7.5cm]{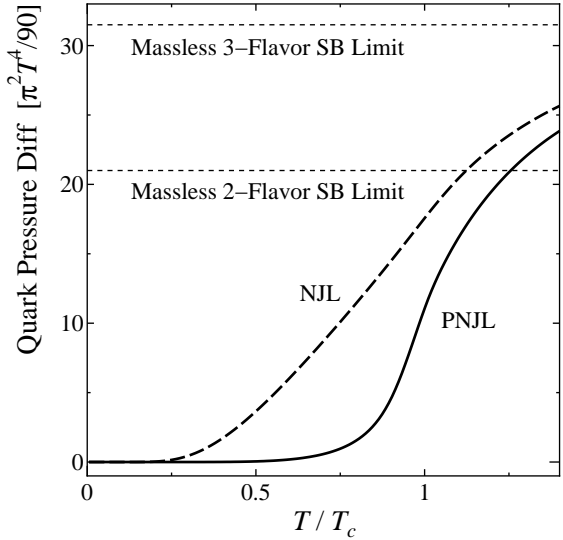}
 \caption{Effective degrees of freedom associated with $2+1$ flavor
 quarks with increasing temperature in the NJL model and in the PNJL
 model.  The critical temperature is $T_c=171.6\MeV$ in the NJL model
 case and $T_c=204.8\MeV$ in the PNJL model case.}
 \label{fig:ed}
\end{figure}

  Let us elucidate how the \textit{effective} confinement is possible
in the model description.  The underlying idea in the PNJL model is
that the group integration (average) with respect to the Polyakov loop
acts as a projection onto the center symmetric state (or the
\textit{canonical ensemble}~\cite{Fukushima:2002bk} with zero
$\mathrm{Z}_3$ charge) if there is no Polyakov loop mean-field.  (See
Eqs.~(\ref{eq:det_lp}) and (\ref{eq:det_la}) and also
Refs.~\cite{Fukushima:2006uv,Megias:2004hj} for details.)  We solve
the four coupled equations~(\ref{eq:gap_eq}) at $T\neq0$ and $\mu=0$,
and plot the quark pressure difference from the zero temperature value
in Fig.~\ref{fig:ed} using the obtained mean-fields.

  In the limit of massless two and three flavors we can count the
fermionic degrees of freedom as
$\nu=(7/8)\cdot3\cdot2\cdot4=21$ and
$\nu=(7/8)\cdot3\cdot3\cdot4=31.5$, respectively.  Because the system
of our interest is quark matter with two light and one heavy flavors,
$\nu$ should take a certain value between 21 and 31.5 at temperature
above $T_c$ where chiral symmetry is restored.  This expectation is
manifest in view of Fig.~\ref{fig:ed} both in the NJL model and in the
PNJL model.  Here we have determined the pseudo-critical temperature
by the location where the temperature derivative,
$\partial\ubu/\partial T$, is largest.  It follows that
$T_c=171.6\MeV$ for the NJL model results and $T_c=204.8\MeV$ for the
PNJL model results (see also Fig.~\ref{fig:chiral_plane}).

  Even in the standard NJL model the effective degrees of freedom go
down as the temperature becomes lower.  This is because quark
excitations are suppressed by the constituent quark mass in the low
temperature side where chiral symmetry is spontaneously broken.  In
reality the system should be mainly composed of a gas of $\pi^0$ and
$\pi^\pm$ below $T_c$ but the $\pi$ mass is $\sim135\MeV$ which is
comparable to the critical temperature.  It is thus expected that we
can neglect the $\pi$ loop corrections in the pressure in the first
approximation.

  We can see from Fig.~\ref{fig:ed} that the NJL model contains too
many (unphysical) quark excitations below $T_c$.  It should be
mentioned that, strictly speaking, too many excitations cannot be
concluded until this comparison and the observation that the PNJL
model is consistent with the lattice results.  These fictitious
excitations diminish only slowly.  It is apparent that the Polyakov
loop projection works efficiently in the PNJL model case.  The
effective degrees of freedom rapidly decrease near $T_c$, that means
that artificial quark excitations are removed by the Polyakov loop
coupling.  Therefore, we can anticipate that the PNJL model should
be more capable to capture realistic thermodynamics than the standard
NJL model especially at temperatures near $T_c$.  Also, because the
Polyakov loop projection affects the quark sector, it is a natural
expectation that the PNJL model would be a more suitable description
than the NJL model in the finite density region where quarks exist
abundantly.

  Finally we shall remark that the separation of the total pressure
into the Polyakov loop and the quark contributions like in
Figs.~\ref{fig:gl} and \ref{fig:ed} does not make sense in the
mean-field approximation.  This is because each of $\ubu$, $\sbs$,
$\ell$, and $\lb$ determined by the gap equations~(\ref{eq:gap_eq})
involve entangled contributions and thus a clear separation is
impossible in any way.


\subsection{Susceptibility}

  In this subsection we clarify how we can evaluate the susceptibility
in respective channels of our interest.  A useful alternative is to be
deduced from the temperature slope, i.e.\ $-\partial\ubu/\partial T$,
$\partial\ell/\partial T$, and so on.  They behave as a function of
$T$ in a similar manner as the susceptibility but the temperature
slope is not really informative more than the order parameter curves
read from Fig.~\ref{fig:chiral_plane}.

\begin{figure}
 \includegraphics[width=8cm]{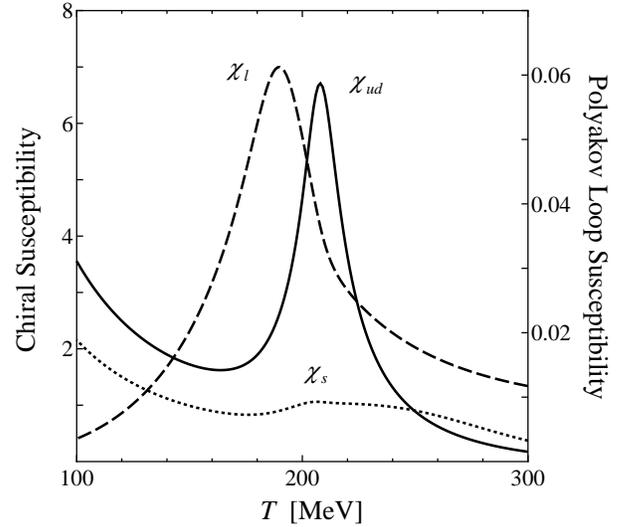}
 \caption{Chiral and Polyakov loop susceptibility at zero density as a
 function of the temperature in the $2+1$ flavor PNJL model.}
 \label{fig:chiral_sus}
\end{figure}

  In order to compute the susceptibility in the mean-field model we
need some caution.  Remembering that the logarithm of the partition
function is $-V\Op/T$, we can give the definition of the dimensionless
susceptibility of our interest as
\begin{align}
 \chi_{ud} &= \frac{1}{4T}
  \frac{\partial^2 (-\Op/T)}{\partial m_{ud}^2} \,,
\label{eq:sus_u}\\
 \chi_s &= \frac{1}{T}
  \frac{\partial^2 (-\Op/T)}{\partial m_s^2} \,,
\label{eq:sus_s}\\
 \chi_{\ell} &= T^3\frac{\partial^2 (-\Op/T)}
  {\partial\eta\,\partial\bar{\eta}} \,,
\label{eq:sus_l}\\
 \chi_q &= \frac{1}{T}
  \frac{\partial^2 (-\Op/T)}{\partial \mu^2} \,,
\end{align}
for the light-quark susceptibility, the heavy-quark susceptibility,
and the Polyakov loop susceptibility, respectively.  We also enumerate
the quark number susceptibility that we will discuss later.  Here we
have inserted the Polyakov loop source $\eta$ and $\bar{\eta}$ in the
potential as $\Omega_{\text{Polyakov}}\to
\Omega_{\text{Polyakov}}-T(\eta\ell+\bar{\eta}\lb)$.  It is crucial to
notice that we have to take the derivative in a way that it hits the
mean-fields also.  That means that we should take
$\partial\ubu/\partial m_{ud}$, $\partial^2\ubu/\partial m_{ud}^2$,
$\partial\sbs/\partial m_{ud}$, $\partial^2\sbs/\partial m_{ud}^2$,
$\partial\ell/\partial m_{ud}$, $\partial^2\ell/\partial m_{ud}^2$,
etc into account to evaluate Eq.~(\ref{eq:sus_u}) for instance.
Otherwise we would miss the loop effect and the mixing to other
channels.

  We can justify this procedure by evaluating the susceptibility in an
independent (and equivalent) method.  By definition, in general, the
susceptibility is to be identified as the inverse of the potential
curvature.  For the purpose to compute the curvature inverse, we
should consider the curvature matrix $C$ whose dimensionless
components are given by
$C_{uu}=T^2\partial^2\Op/\partial\ubu^2$,
$C_{us}=T^2\partial^2\Op/\partial\ubu\partial\sbs$,
$C_{u\ell}=T^{-1}\partial^2\Op/\partial\ubu\partial\ell$,
$C_{\ell\lb}=T^{-4}\partial^2\Op/\partial\ell\partial\lb$, and so on.
In the present case $C$ is a $4\times 4$ matrix.  Then the diagonal
components of $C^{-1}$ give the susceptibility which is an involved
expression in terms of $C_{uu}$, $C_{us}$, $C_{u\ell}$, etc.  Roughly
speaking, the diagonal part, $C_{uu}^{-1}$, $C_{ss}^{-1}$,
$C_{\ell\lb}^{-1}$, corresponds to soft-mode propagators and the
off-diagonal part, $C_{us}$, $C_{u\ell}$, $C_{\ell\ell}$,
$C_{\lb\lb}$, and so on, corresponds to mixing vertices.  It is
immediate to make sure that $C^{-1}$ certainly leads to exactly the
same results as obtained from Eqs.~(\ref{eq:sus_u}), (\ref{eq:sus_s}),
and (\ref{eq:sus_l}).  This matrix method has an advantage in giving
us the mixing angle between each mode.

  As we can notice from Fig.~\ref{fig:chiral_sus} showing the
susceptibility as a function of $T$, two crossovers associated with
$\ubu$ and $\ell$ are located close to each other but do not coincide
precisely.  As long as we treat the chiral condensate and the Polyakov
loop as independent variables as in the PNJL model, two crossovers
attract each other to some extent but have a short ``repulsion.''
Within this kind of model approach it is hence hard to explain the
complete coincidence without fine tuning.

  One interesting strategy is not to explain the locking but to build
a new model based on the complete locking of chiral restoration and
deconfinement.  As discussed in Ref.~\cite{hatta}, most of lattice
results support the idea that there is only one order parameter field
$\phi$ that is a mixture of the $\sigma$ meson and the electric
glueball (Polyakov loop).  Then, we could make a model with the chiral
condensate given by $\ubu\propto \phi\cos\theta$ and the Polyakov loop
by $\ell\propto \phi\sin\theta$ with some potential energy for the
mixing angle $\theta$ between them.  The work along this direction is
under progress~\cite{future}.


\subsection{Quark Number Susceptibility}

\begin{figure}
 \includegraphics[width=7.5cm]{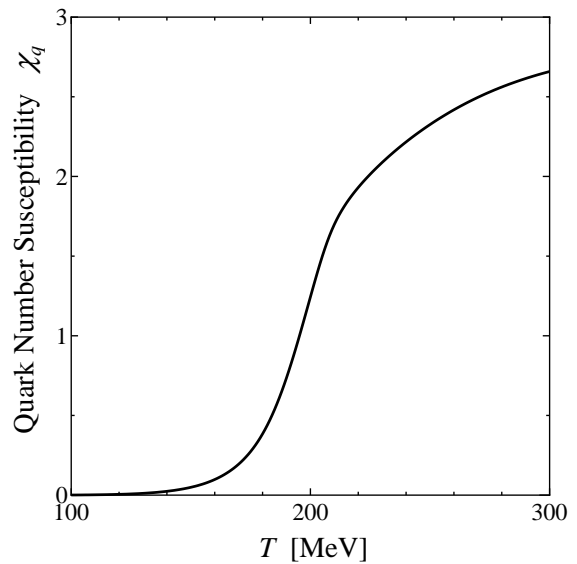}
 \caption{Quark number susceptibility calculated in the PNJL model for
 $2+1$ flavor quark matter at zero density.}
 \label{fig:quark_num}
\end{figure}

\begin{figure}
 \includegraphics[width=7.5cm]{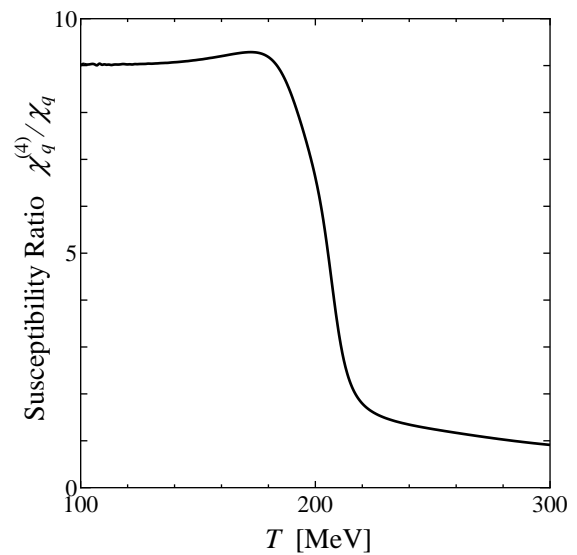}
 \caption{Ratio of the fourth derivative to the quark number
 susceptibility calculated in the PNJL model for $2+1$ flavor quark
 matter at zero density.}
 \label{fig:quark_num_r}
\end{figure}

  It is difficult to probe physical observables sensitive to the
chiral and Polyakov loop susceptibility directly in experiments.  In
fact, it is impossible to count the number fluctuation of the $\sigma$
meson and the glueball which eventually decay to the lightest $\pi$
meson.  From the experimental point of view the quark number
susceptibility should be a better measure because the quark number is
a conserved quantity.  The fluctuation in the baryon multiplicity
would be directly related to the quark number susceptibility,
$\chi_q$~\cite{Kunihiro:1991qu,Hatsuda:1994pi,Hatta:2002sj}.  Also in
Refs.~\cite{Ratti:2007jf,Sasaki:2006ww} $\chi_q$ has been evaluated
and discussed in the two-flavor PNJL model.  Actually the PNJL model
can reproduce $\chi_q$ measured on the lattice in the two-flavor case
as beautifully illustrated in Ref.~\cite{Ratti:2007jf}.

  We plot our results in the $2+1$ flavor case in
Fig.~\ref{fig:quark_num}.  We can see, as expected, that the $2+1$
flavor quark matter yields $\chi_q$ greater than the two-flavor case
shown in Ref.~\cite{Ratti:2007jf}.  In the chiral limit $\chi_q$ would
scale as $N_f^2$, and thus the three-flavor value should be
$3^2/2^2=2.25$ times greater than the two-flavor value.  Because
$s$-quarks are massive in reality, this scale factor should become
smaller.  Let us choose one temperature to take an example for
comparison.  At the temperature $T=1.5T_c\simeq300\MeV$,
Fig.~\ref{fig:quark_num} reads around $2.7$, while the two-flavor
value is around $1.5$, which leads to the ratio $1.8$.  This seems to
be a reasonable number.

  It is interesting to define the following
quantity~\cite{Ejiri:2005wq};
\begin{equation}
 \chi^{(4)}_q = T \frac{\partial^4(-\Op/T)}{\partial\mu^4} \,,
\end{equation}
and take the ratio to the quark number susceptibility.  This ratio,
$\chi^{(4)}_q/\chi_q$, counts the number squared of quark content
inside thermally excited particles carrying baryon number.  Therefore,
if quarks are liberated in the high temperature region,
$\chi^{(4)}_q/\chi_q\simeq1^2$ should follow, whereas the low
temperature side should results in $\chi^{(4)}_q/\chi_q\simeq3^2$
because of confinement.  This is actually a general feature in the
quasi-particle picture and attributed to the Boltzmann factor in the
free fermionic partition function.

  Figure~\ref{fig:quark_num_r} shows this susceptibility ratio
obtained in the $2+1$ flavor PNJL model.  We see that the behavior
perfectly fits what is expected.  A short conclusion that we should
learn from this analysis is that $\chi^{(4)}_q/\chi_q$ signifies the
quark number but does not tell us whether the thermally excited
particle is a confined nucleon or a set of three quarks.  The latter
is the case in the PNJL model.


\subsection{More on Thermodynamics}

\begin{figure}
 \includegraphics[width=7.5cm]{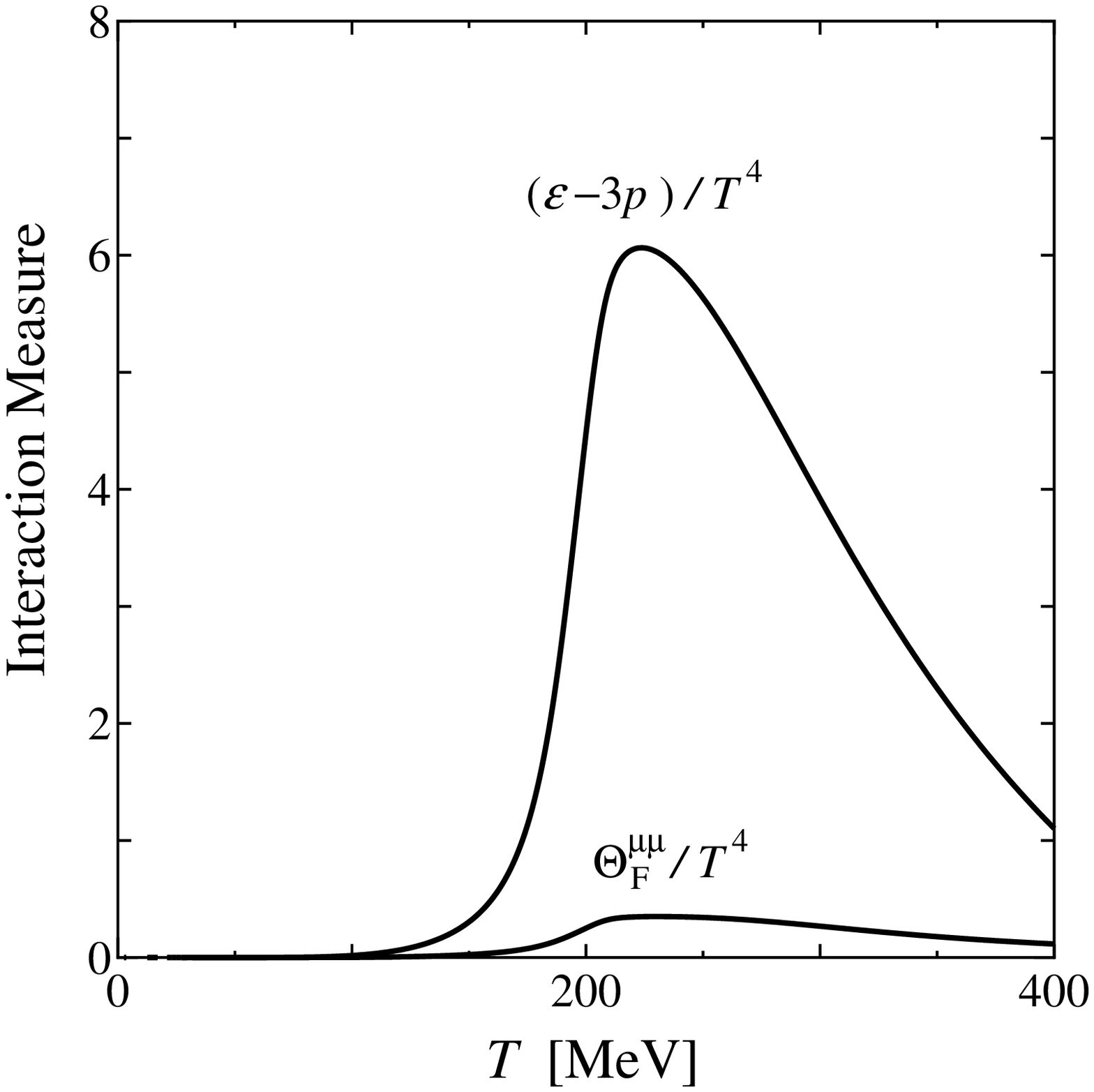}
 \caption{Plot for the ``interaction measure,'' i.e.\
 $(\epsilon-3p)/T^4$, and the fermion contribution,
 $\Theta^{\mu\mu}_{\text{F}}=[2m_{ud}(\ubu-\ubu_0)+m_s(\sbs-\sbs_0)]$.}
 \label{fig:anomaly}
\end{figure}

\begin{figure}
 \includegraphics[width=7.5cm]{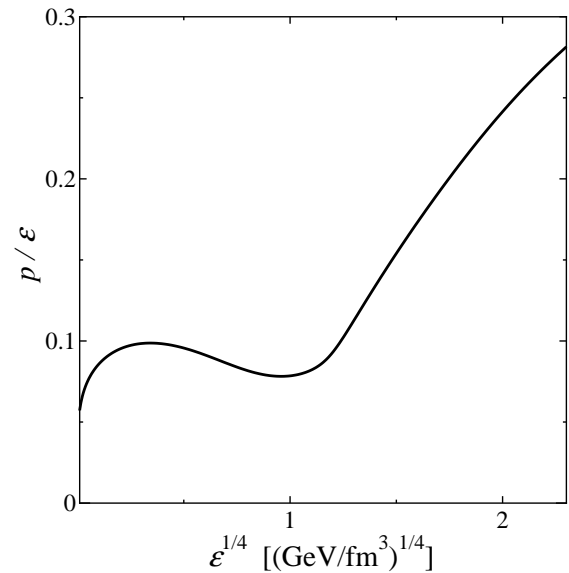}
 \caption{Plot for $p/\epsilon$ as a function of $\epsilon^{1/4}$.}
 \label{fig:velocity}
\end{figure}

  Before proceeding into the finite density inquiry, we shall
exemplify the success of the PNJL model by two more thermodynamic
quantities.

  The trace of the energy momentum tensor is vanishing at the
classical level when theory has no mass scale.  We know that QCD in
the chiral limit is scale invariant, which means that the trace of the
energy momentum tensor in massless QCD is zero unless quantum
corrections are taken into account.  The QCD scale
$\Lambda_{\text{QCD}}$ arises from the dimensional transmutation due
to the trace anomaly at the quantum level.

  In thermodynamics the traceless of the energy momentum tensor
without mass gap means $\epsilon-3p=0$ where $p$ is the pressure given
by $-\Op$ and $\epsilon$ is the internal energy given by
$T^2\partial(p/T)/\partial T$.  It is straightforward to evaluate
$\epsilon-3p$  or the ``interaction measure'' from $\Op$ in our model
study.

  We show the model results in Fig.~\ref{fig:anomaly}.  The gross
structure with a peak around $T_c$ is in nice agreement with the
recent lattice data (see Fig.~4 in Ref.~\cite{Cheng:2007jq}).  The
peak height in the interaction measure is not as large as that in
Ref.~\cite{Cheng:2007jq}, which is partly because of the finite number
of $N_\tau$ on the lattice and partly because of the smaller fermion
contribution, $\Theta^{\mu\mu}_{\text{F}}$, in our calculation.  We
present the results for
$\Theta^{\mu\mu}_{\text{F}}=2m_{ud}(\ubu-\ubu_0)+m_s(\sbs-\sbs_0)$
also in Fig.~\ref{fig:anomaly}.  We see that our results are
significantly smaller than the results shown in Fig.~5 in
Ref.~\cite{Cheng:2007jq}.  This is because the quark mass is
different;  the $\pi$ mass in Ref.~\cite{Cheng:2007jq} is still around
$220\MeV$, while we choose $m_{ud}$ to yield the realistic $\pi$
mass.

  We should be aware that the interaction measure,
$(\epsilon-3p)/T^4$, has only little to do with the trace anomaly in
the PNJL model study.  We have model inputs with mass dimension, that
is, the cutoff $\Lambda$.  (There are four more dimensional
parameters, $\gs$, $\gd$, $a$, and $b$ but they can be all
dimensionless in unit of $\Lambda$.)

  As a matter of fact, the peak structure is rather generic regardless
of any specific model.  One can understand this from the thermodynamic
relation,
\begin{equation}
 \frac{\epsilon-3p}{T^4} = T\frac{\partial}{\partial T}\Bigl(
  \frac{p}{T}\cdot\frac{1}{T^3}\Bigr) \,.
\end{equation}
The right-hand side is the temperature derivative of $p/T^4$ where
$p/T^4$ naively counts the effective degrees of freedom $\nu$ as
plotted in Figs.~\ref{fig:gl} and \ref{fig:ed}.  Therefore, so-called
the trace anomaly, $\epsilon-3p$, signifies how quickly $\nu$ grows up
as $T$ increases.  The pseudo-critical temperature is, by definition,
where $\nu$ starts getting larger, and eventually $\nu$ is saturated
to the total number of particle species at high temperature.  As a
result the peak shape is inevitable associated with crossover
behavior.  It is not quite surprising in this sense that the PNJL
model can mimic the trace anomaly in hot QCD around $T_c$.

  In other words, it is the relation between $\epsilon-3p$ and the
gluon condensate that is a non-trivial consequence from the trace
anomaly.  The interaction measure, $(\epsilon-3p)/T^4$, is governed
not by the anomaly but by the thermodynamics which determines the
gluon condensate in turn.

  Now that we have come by the pressure and the internal energy, we
can infer the sound velocity.  Although the velocity of sound is given
by $c_s^2=\rmd p/\rmd \epsilon$, the ratio $p/\epsilon$ can
approximate it in the high temperature limit.  To compare our results
to the available lattice data, we plot $p/\epsilon$ as a function of
$\epsilon^{1/4}$ in Fig.~\ref{fig:velocity}, which agrees quite well
with Fig.~9 in Ref.~\cite{Cheng:2007jq}.  We remark that the sound
velocity has been investigated by means of the two-flavor PNJL model
also in Ref.~\cite{Ghosh:2006qh} where both of $c_s^2$ and
$p/\epsilon$ are presented.


\section{FINITE DENSITY RESULTS}

  By adding one more axis in the direction of quark chemical potential
we can investigate the order parameter behavior and the phase
structure in wider perspective.  In this work we limit ourselves to
the chiral and deconfinement phase transitions and do not take account
of the diquark condensation that plays an essential role in the color
superconducting
phase~\cite{Roessner:2006xn,Ciminale:2007ei,Abuki:2008ht,future2}.


\subsection{Chiral Phase Transition}

\begin{figure}
 \includegraphics[width=7.5cm]{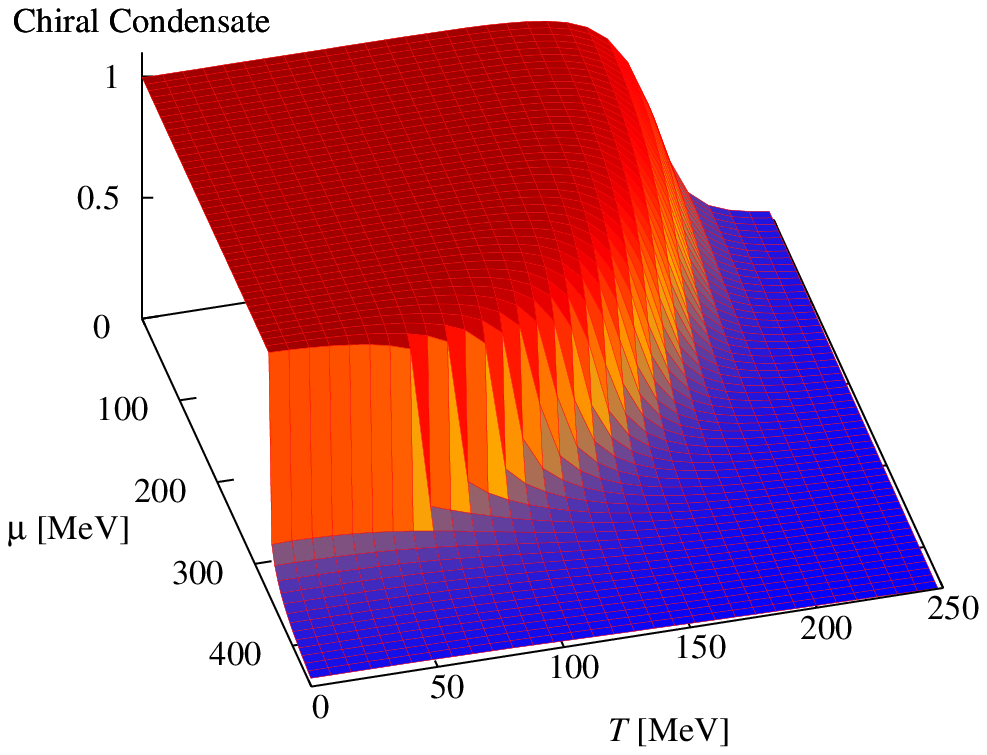}
 \caption{Normalized light-quark chiral condensate $\ubu/\ubu_0$ in
 the $\mu$-$T$ plane, where $\ubu_0$ is the chiral condensate at
 $T=\mu=0$.}
 \label{fig:chiral}
\end{figure}

  Figure~\ref{fig:chiral} is a 3D plot for $\ubu$ as a function of $T$
and $\mu$.  We see that there is a discontinuity in the low
temperature and high density region, while the high temperature and
low density region has continuous crossover.  Therefore the phase
diagram has an end-point of the first-order phase boundary, that is
called the QCD critical point.  We remark that in the present
parameter set the constituent quark masses turn out to be
\begin{equation}
 M_{ud} = 336\MeV\,,\quad M_s = 528\MeV\,,
\end{equation}
and the first-order phase transition is located at $\mu=345\MeV$ when
$T=0$, which is slightly above the light quark mass.  This is a
general feature to be explained intuitively.  First, let us focus on
the region at $\mu<345\MeV$ where the system does not have any
discontinuous transition along the $T=0$ density axis.  We can still
locate the point where a non-vanishing baryon density appears at
$\mu=M_{ud}$, that is a sort of continuous phase transition from the
empty vacuum to degenerate quark matter.  Next, once quark matter is
concerned at $\mu\simeq345\MeV$, the pressure of cold quark matter at
a fixed value of $\mu$ becomes smaller for larger quark mass;  for
instance $p\propto\mu^4$ for massless quarks and
$p\propto(\mu^2-M^2)^2$ for massive quarks.  Thus, the kinetic energy
favors lighter quark matter, that is, the chiral symmetric phase.  The
condensation energy gives a negative contribution to the pressure,
that means that the chiral symmetric phase where the condensation
energy is smaller is energetically favorable again.  In this way, one
can expect that, as soon as the quark number density becomes
substantial with $\mu$ going above $M_{ud}$, the system tends to
undergo a phase transition to the chiral symmetric phase.


\subsection{Polyakov Loop}

\begin{figure}
 \includegraphics[width=8cm]{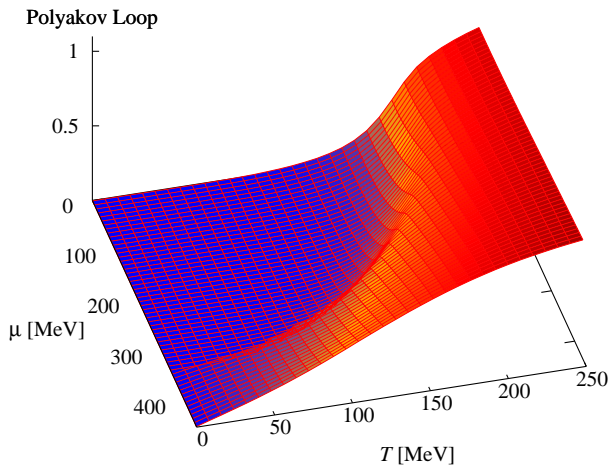}
 \caption{Polyakov loop $\ell$ in the $\mu$-$T$ plane.}
 \label{fig:pol}
\end{figure}

  It is interesting to see what happens in the Polyakov loop behavior
on the $\mu$-$T$ plane.  One may well anticipate that the coincidence
of chiral restoration and deconfinement should persist in the finite
density region.  This expectation is partially true but too na\"{i}ve.
We shall discuss the appropriate physical interpretation in what
follows below.

  We plot the Polyakov loop $\ell$ in the $\mu$-$T$ plane in
Fig.~\ref{fig:pol}.  It should be mentioned that we do not make
another plot for $\lb$, for $\lb$ has a qualitatively same functional
shape as $\ell$ with small quantitative difference.

  From the comparison between the chiral condensate displayed in
Fig.~\ref{fig:chiral} and the Polyakov loop in Fig.~\ref{fig:pol}, we
can readily perceive that two crossovers are linked in the entire
region on the $\mu$-$T$ plane.  For instance, we have already
confirmed that two crossovers are simultaneous indeed at zero density
in Fig.~\ref{fig:chiral_plane}, and we can find a first-order phase
transition along the density axis at low temperature whose location
is exactly the same in Figs.~\ref{fig:chiral} and \ref{fig:pol}.  The
locking of chiral restoration and deconfinement remains at finite
density in this sense.

  It would be misleading, however, to dive into a conclusion that two
phenomena of chiral restoration and deconfinement simultaneously take
place in the high density region.  In view of the Polyakov loop
behavior at low temperatures, in fact, the discontinuous jump is tiny
and the expectation value of the Polyakov loop stays vanishingly small
even at $\mu>345\MeV$ where chiral symmetry is restored.  Therefore,
the discontinuous jump in the Polyakov loop signifies a first-order
phase transition from nearly confined matter ($\ell\simeq0$) with
chiral symmetry breaking ($\ubu\neq0$) to nearly confined matter
($\ell\simeq0$) with chiral symmetry restoration ($\ubu\simeq0$).

  It is an interesting question how the Polyakov loop behaves such
differently from the chiral condensates in the region of low
temperature and high density.  This is because center symmetry is not
broken at zero temperature even in the presence of dynamical quarks,
and therefore, the expectation value of the Polyakov loop must stay
vanishing.  The reason for preserved center symmetry is to be
understood intuitively as follows;  when the quark density is
specified by a certain chemical potential, each energy level is
occupied by a quark up to the Fermi surface.  Because quarks have
color degeneracy, red, green, and blue quarks always sit on the same
energy level, which makes a color singlet.  One can easily see this
really happening in the model from the Dirac determinant given in
Eqs.~(\ref{eq:det_lp}) and (\ref{eq:det_la}).  That is, when
$\mu>\varepsilon$ we only have the second term out of the whole
particle contribution,
\begin{equation}
 1+\rme^{3|\varepsilon-\mu|/T}+3\,\ell\,\rme^{|\varepsilon-\mu|/T}
  +3\,\lb\,\rme^{2|\varepsilon-\mu|/T} \,,
\label{eq:det_p}
\end{equation}
that is exponentially dominant for large $|\varepsilon-\mu|/T$.  This
second term, $\rme^{3|\varepsilon-\mu|/T}$, actually represents the
three-quark occupation which does not couple $\ell$ and thus not break
center symmetry.  The third and fourth terms are one-quark and
two-quark (which is equivalent to one anti-quark in color)
contributions with non-trivial (non-singlet) color.  Consequently, the
Polyakov loop is insensitive to whether the quark degrees of freedom
are present in the system or not.  Strictly speaking, the PNJL model
cannot deal with confinement, namely, nucleon wavefunctions as a bound
state out of three quarks, but still, the low temperature region
always has a signature of confinement ($\ell\simeq0$) with respect to
quarks.  We would stress that this quarky confined phase is not a
model artifact but physical one.  We propose that this phase should be
identified as the \textit{quarkyonic phase} discussed in
Ref.~\cite{McLerran:2007qj}.

\begin{figure}
 \includegraphics[width=8cm]{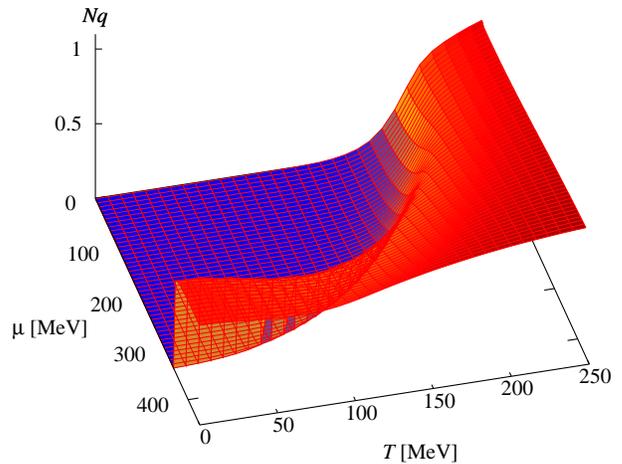}
 \caption{Quark number density normalized by the free massless quark
 density, $N_c N_f (\mu^3/3\pi^2+T^2\mu/3)$, in the $\mu$-$T$ plane.}
 \label{fig:nq}
\end{figure}

  One important suggestion emphasized in Ref.~\cite{McLerran:2007qj}
is that the baryon or quark number density serves as an order
parameter to tell the quarkyonic phase from the hadronic phase.  We
have then calculated the quark number density in the $\mu$-$T$ plane
to make a plot of Fig.~\ref{fig:nq}.  The quark number density is
readily available from
\begin{equation}
 n_q = -\frac{\partial\Op}{\partial \mu} \,.
\label{eq:nq}
\end{equation}
In order to visualize in a sensible manner on the 3D plot, we
normalize $n_q$ by the free massless quark density given by
$N_c N_f(\mu^3/3\pi^2+T^2\mu/3)$, that is, in Fig.~\ref{fig:nq} we
plot,
\begin{equation}
 N_q = \frac{n_q}{3\mu^3/\pi^2 + 3T^2\mu} \,,
\end{equation}
which should take a value from zero to unity.

  We can clearly confirm that the quark degrees of freedom are
relevant (i.e.\ $N_q\sim\mathcal{O}(1)$) even in the region at
$T\simeq0$ and $\mu>M_{ud}$ where $\ell\simeq0$.  This gives another
evidence for our identification to the quarkyonic phase.

  It is a non-trivial finding from the present PNJL model study that
$N_q$ surely behaves as an order parameter and locates the crossover
point that coincides the chiral phase transition point.  This
coincidence is apparent at a glance of Figs.~\ref{fig:chiral} and
\ref{fig:nq}.


\subsection{Phase Diagram}

  We now explore the phase diagram in the $\mu$-$T$ plane by the
cross-section of Figs.~\ref{fig:chiral}, \ref{fig:pol}, and
\ref{fig:nq} at a certain height in the vertical axis.  As we
discussed, in the high density region in particular, the
susceptibility peak position does not make much sense, but the
magnitude of the order parameter is a more suitable quantity to probe
the physical state of matter.  For example, the Polyakov loop
susceptibility diverges at the critical end-point as well as the
chiral susceptibility, but it does not result from the deconfinement
transition but from the mixing to the chiral dynamics.  Therefore we
define the pseudo-critical temperature for $u$-quark chiral
restoration by the condition;
\begin{equation}
 \frac{\ubu}{\ubu_0}\Biggr|_{T=T_u(\mu)} \!\!\!= \frac{1}{2} \,.
\end{equation}
In the same way we can define the pseudo-critical temperature for
$s$-quark chiral restoration by
\begin{equation}
 \frac{\sbs}{\sbs_0}\Biggr|_{T=T_s(\mu)} \!\!\!= \frac{1}{2} \,.
\end{equation}
Also, we can define the pseudo-critical temperature for deconfinement
by
\begin{equation}
 \ell\;\Bigr|_{T=T_\ell(\mu)} = \frac{1}{2} \,.
\label{eq:Td}
\end{equation}

\begin{figure}
 \includegraphics[width=7.5cm]{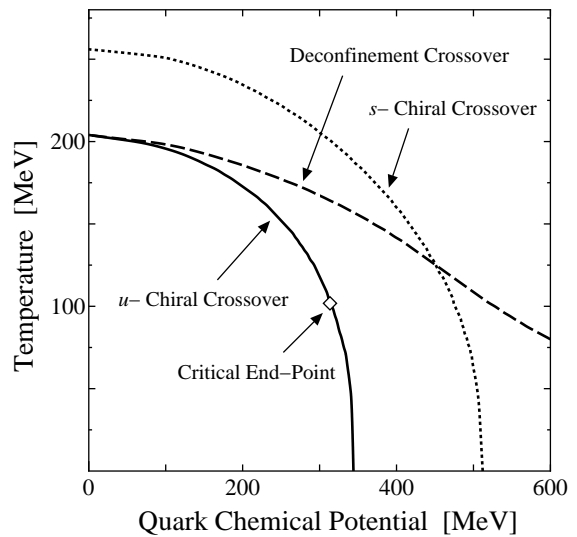}
 \caption{Phase diagram characterized by three quantities, namely, the
 $u$-quark chiral condensate, the $s$-quark chiral condensate, and the
 Polyakov loop.}
 \label{fig:phase}
\end{figure}

  Then, we can draw three distinct curves by $T=T_u(\mu)$, $T_s(\mu)$,
$T_\ell(\mu)$ on the $\mu$-$T$ phase diagram.  The PNJL model
prediction is shown in Fig.~\ref{fig:phase}.  The solid curve
represents $T=T_u(\mu)$ which is crossover in the low density region
and turns a first-order phase transition in the high density region
accompanied by a critical end-point.  We note that $N_q$ is nearly
zero inside this solid curve.  Because of explicit symmetry breaking
by $m_s\neq0$, the $T=T_s(\mu)$ boundary is located at higher $T$ and
$\mu$ shown by the dotted curve in Fig.~\ref{fig:phase}.  Of course,
strictly speaking, chiral symmetry or even $\mathrm{SU_V}(3)$ symmetry
is not restored at any temperature or density, but $\sbs$ can decrease
up to a half of $\sbs_0$ smoothly.  Actually this boundary hits $T=0$
and $\mu=512\MeV$ which is not far from the constituent $s$-quark
mass.  In any case, the boundary by the dotted curve does not have a
strong meaning because the change in $\sbs$ as a function of $T$ is
only gradual.  Nevertheless, the region bounded by
$T_u(\mu)<T<T_s(\mu)$ is interesting.  This is because the
$\mathrm{SU_V}(3)$ symmetry breaking becomes enhanced further in this
region by chiral restoration for $u$-quarks and $d$-quarks but not for
$s$-quarks~\cite{kunihiro}.

  It is surprising that the deconfinement crossover defined by the
condition~(\ref{eq:Td}) goes away from the chiral phase transition at
finite density as indicated by the dashed curve which represents the
$T=T_\ell(\mu)$ curve.  As we have discussed, the Polyakov loop
expectation value is always vanishing at zero temperature, and thus,
this $T=T_\ell(\mu)$ curve never hits the horizontal axis at $T=0$.
The region surrounded by $T_u(\mu)<T<T_\ell(\mu)$ is what should be
called the quarkyonic phase embodied in the PNJL model.

  As a final remark in this section we refer to the similar results
presented in Figs.~16 and 17 in Ref.~\cite{Sasaki:2006ww} and the
similar physical picture to the quarkyonic phase discussed in a
different context in Ref.~\cite{Glozman:2007tv}.


\section{QCD CRITICAL END-POINT}

  We have already mentioned on the QCD critical end-point in the
explanation of Fig.~\ref{fig:phase}.  The rest of this paper will be
devoted to physics related to the QCD critical point.  First of all,
it is instructive to elucidate how the location of the critical point
moves by the effect of the Polyakov loop.  In the three-flavor NJL
model with the Hatsuda-Kunihiro parameters, the location of the
critical point is found to be
\begin{equation}
 (T_E,\, \mu_E) = (48\MeV,\, 324\MeV) \,,
\end{equation}
in the three-flavor NJL model.  The location is almost the same as in
the two-flavor case.  [See Ref.~\cite{Kashiwa:2007hw} for a summary
table and also Ref.~\cite{Costa:2008gr}.]  The model dependence is
nicely compiled also on Fig.~4 in Ref.~\cite{Stephanov:2007fk}.  In
the three-flavor PNJL model the location goes up along the temperature
to
\begin{equation}
 (T_E,\, \mu_E) = (102\MeV,\, 313\MeV) \,,
\end{equation}
in the present parameter set.  This value is close to the two-flavor
PNJL location first reported in my paper~\cite{Fukushima:2003fw}.  The
reason why the critical point moves toward higher temperature is that
the artificial quark excitation at finite temperature and density is
suppressed by the Polyakov loop average as exhibited in
Fig.~\ref{fig:ed} and also discussed around Eq.~(\ref{eq:det_p}).

  The question is to what extent we can trust the model prediction for
the location of the QCD critical point or even its existence.  In what
follows we will discuss the dependence on ambiguous model parameters.
So far, it is quite difficult to make any robust statement about the
QCD critical point from model studies, that is our short conclusion.


\subsection{Divergent Susceptibility}

  Before addressing the theoretical uncertainty on the QCD critical
end-point, we will begin with standard arguments, that is, physical
implication from the assumption that the QCD phase diagram holds a
critical end-point.

\begin{figure}
 \includegraphics[width=8cm]{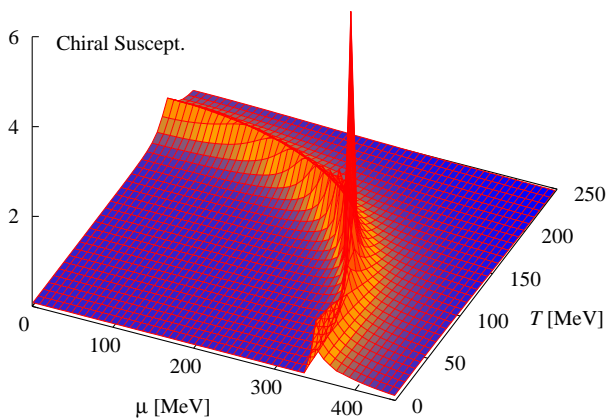}
 \caption{3D plot for the light-quark chiral susceptibility
 $\chi_{ud}$ multiplied by $(T/\Lambda)^2$ in the $\mu$-$T$ plane.}
 \label{fig:sus}
\end{figure}

\begin{figure}
 \includegraphics[width=8cm]{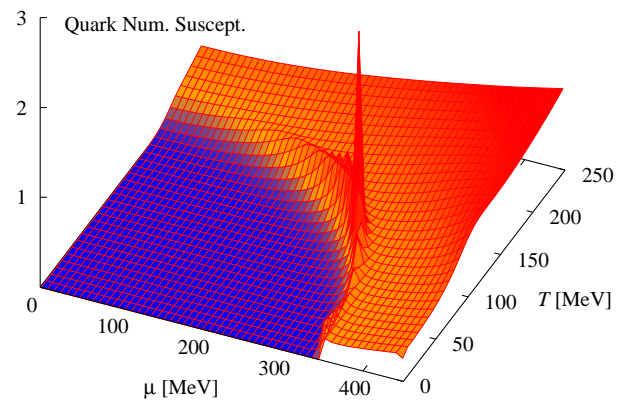}
 \caption{3D plot for the quark number susceptibility $\chi_q$
 multiplied by $(T/\Lambda)^2$ in the $\mu$-$T$ plane.}
 \label{fig:qsus}
\end{figure}

  The importance of the QCD critical point lies in the fact that it is
an exact second-order phase transition point.  Therefore the
susceptibility diverges right at the end-point.  Originally divergent
growth in the chiral susceptibility $\chi_u$ has been paid
attention~\cite{Stephanov:1998dy} which might lead to furious
fluctuations in the $\sigma$ channel and thus $\pi$ fluctuations
through the $\sigma\leftrightarrow 2\pi$ coupling.  We have made a 3D
plot in Fig.~\ref{fig:sus} to show the $u$-quark chiral susceptibility
multiplied by $(T/\Lambda)^2$, i.e.\
$-\frac{1}{4}\Lambda^{-2}\partial^2\Op/\partial m_{ud}^2$ in the
$\mu$-$T$ plane.  We notice that the susceptibility has a singularity
at the critical point.

  A physical quantity of more experimental interest is the quark
number susceptibility.  We plot $\chi_q$ multiplied by $(T/\Lambda)^2$
in Fig.~\ref{fig:qsus}, that is,
$-\Lambda^{-2}\partial^2\Op/\partial\mu^2$ in the $\mu$-$T$ plane.
Figure~\ref{fig:qsus} shows a singularity at the QCD critical point
which should translate into event-by-event fluctuations of baryon
multiplicity.  The global shape is just similar to that of the chiral
susceptibility.  It is, however, different that the quark number
susceptibility gets non-vanishing in the high temperature or density
region whose behavior is closely linked to the quark number density in
Fig.~\ref{fig:nq}.


\subsection{Columbia Diagram}

  What we will reveal particularly in this work is the robustness of
the existence of the critical end-point, which is in part motivated by
the lattice suggestion~\cite{deForcrand:2006pv}.  We can disclose
another aspect of the phase diagram in the plane of the light and
heavy quark masses~\cite{Brown:1990ev}.  Such a phase diagram is
sometimes referred to as the ``Columbia Diagram.''

\begin{figure}
 \includegraphics[width=8cm]{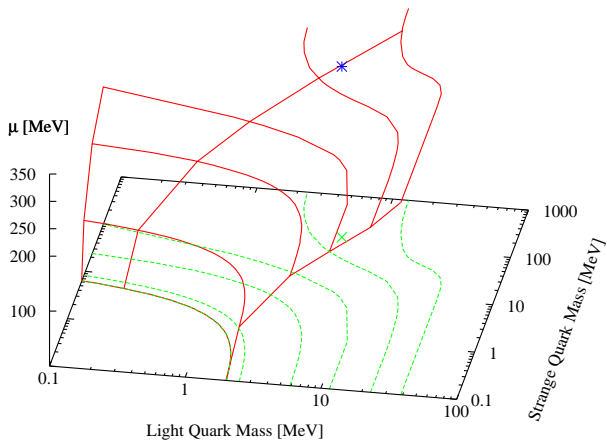}
 \caption{Boundaries of the first-order phase transition region as a
 function of the quark masses at $\mu=0$, $100$, $200$,
 $250$, $300$, $350\MeV$ from the bottom to the top.}
 \label{fig:phase_mm}
\end{figure}

  The PNJL model results are summarized in Fig.~\ref{fig:phase_mm}.
Each curve represents the boundary between the first-order phase
transition to the crossover.  For $m_{ud}$ and $m_s$ below the curve,
the chiral phase transition at finite $T$ is of first-order, and
otherwise, it is crossover.  We draw a line $m_s/m_{ud}=24.67$ which
crosses the physical point, and add two lines at $m_{ud}=0$ and
$m_s=0$, respectively, for the eye guide.

  This plot poses us a problem in the model study based on the
NJL-type description.  It is that the first-order phase transition
region at $\mu=0$ is substantially smaller than what has been observed
in the lattice QCD simulations.  In the $m_s=0$ case, for instance,
the critical value of the light-quark mass is $m_u=2.1\MeV$ in this
work, and on the $m_{ud}=0$ axis, the critical strange quark mass is
$m_s=8.8\MeV$, which are smaller by one order of magnitude at least as
compared to the lattice empirical value~\cite{deForcrand:2006pv}.
This fact implies that the first-order phase transition with massless
three flavors is presumably weaker in the NJL model than realistic.
Then, the critical end-point at physical quark mass cannot avoid being
far away from zero density.  That is, the density has to increase
significantly until the boundary eventually hits the physical mass
point.  This is why the NJL-type model has common tendency to lead to
the critical end-point at relatively high density above
$\mu\sim300\MeV$.

  Because the PNJL model predicts the QCD critical point, the
first-order transition region expands with increasing $\mu$ as shown
in Fig.~\ref{fig:phase_mm}.  The boundary surface is thus standard but
not quite consistent with the recent lattice
observation~\cite{deForcrand:2006pv}.  This is problematic to the
model study if the lattice results are correct.  The model study has,
however, unknown factors which could make a drastic change in the
order.  Here, we will point out two major effects;  one is the $\UA$
anomaly reduction in a medium and the other is the induced
vector-channel interaction.


\subsubsection{Anomaly strength}

  We have already noted that the first-order transition region on the
Columbia diagram obtained in the PNJL model is significantly smaller
than the lattice results.  One likely explanation for this is that the
't~Hooft (six-quark) coupling constant, $\gd$, is weaker in the NJL
model estimate than realistic because of cutoff artifact.  It should
be mentioned that the value of $\gd$ is fixed to reproduce the $\eta'$
mass, which is as large as $957.5\MeV$ and is greater than the cutoff
$\Lambda=631.4\MeV$.  It is not unlikely that the strength of
$\gd$ has been underestimated to reproduce such a large mass in this
cutoff model.

\begin{figure}
 \includegraphics[width=7.5cm]{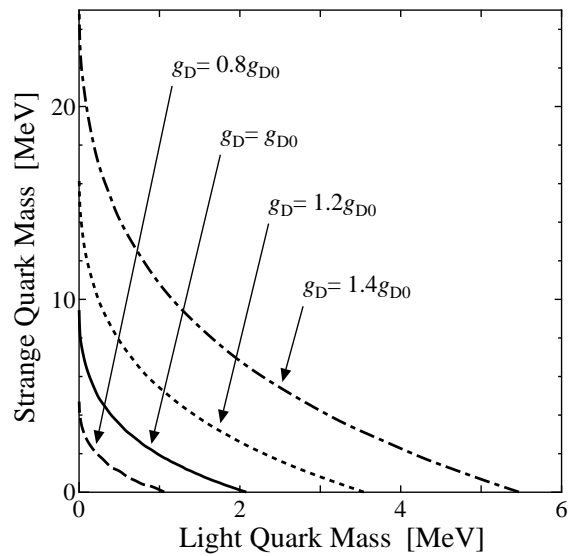}
 \caption{First-order transition boundary depending on the strength of
 the 't~Hooft coupling $\gd$, where ${\gd}_0$ is a value fixed in the
 vacuum in the Hatsuda-Kunihiro parameter set.}
 \label{fig:gd}
\end{figure}

  Figure~\ref{fig:gd} shows the first-order transition boundary on the
$m_{ud}$-$m_s$ plane.  Here ${\gd}_0$ denotes the standard value in
the Hatsuda-Kunihiro parameter set.  Because ${\gd}_0$ has been fixed
in the vacuum, $\gd$ in a hot and dense medium might take a different
(presumably smaller) value.  As we anticipated, the first-order
region becomes wider with larger $\gd$ and narrower with smaller
$\gd$.  It should be noted that the plot is made in the linear scale
in Fig.~\ref{fig:gd}, while the scale is logarithmic in
Fig.~\ref{fig:phase_mm}.

\begin{figure}
 \includegraphics[width=7.5cm]{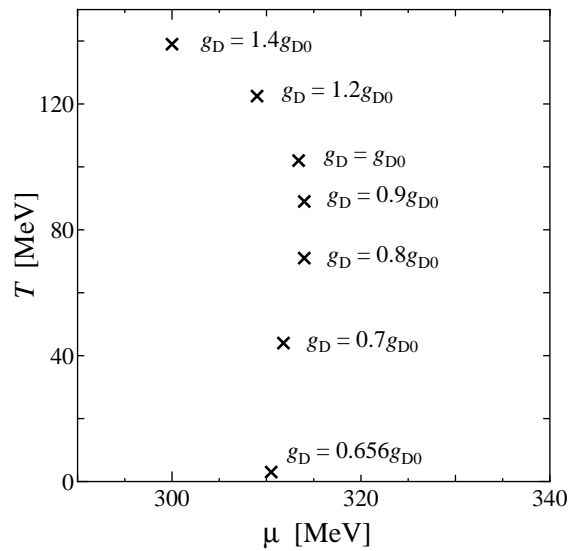}
 \caption{Dependence of the critical point location on the strength of
 the 't~Hooft coupling constant.}
 \label{fig:nocep}
\end{figure}

  One can then expect that the QCD critical point should move
accordingly as $\gd$ changes.  We show the location of the QCD
critical point for various values of $\gd$ in Fig.~\ref{fig:nocep}.
We can learn two lessons from this figure:  One is that the QCD
critical point can be located at higher temperature and lower density
if $\gd$ is underestimated in the NJL model study due to too heavy
$\eta'$ out of model reliability.  The other is in a sense opposite to
the first one.  The QCD critical point might be absent from the QCD
phase diagram if $\gd$ is suppressed by the effective $\UA$
restoration at high density.  Actually, only $35\%$ reduction is
enough to make the QCD critical point disappear from the phase
diagram.  If the suppression is exponential like
$\gd(\mu)=e^{-\mu^2/\mu_0^2}{\gd}_0$~\cite{Hatsuda:1994pi}, $35\%$
reduction is within a reasonable reach.

  Then, one could change the scenario in Fig.~\ref{fig:phase_mm} by
introducing a $\mu$-dependent value for $\gd$.  For instance, if one
assumes an exponential ansatz, $\gd(\mu)=e^{-\mu^2/\mu_0^2}{\gd}_0$,
one could find some $\mu_0$ that produces a boundary surface with
bending behavior that the first-order transition region shrinks with
increasing $\mu$.


\subsubsection{Vector-channel interaction}

  It is not only the $\UA$ anomaly term but also the vector-channel
interaction term in Eq.~(\ref{eq:vec}) that can affect the location of
the QCD critical point.  We remark that $\Lag_V$ does not break chiral
symmetry at all, and besides, the zeroth component corresponds to the
density operator $(\psi^\dagger\psi)^2$.  Therefore, it is conceivable
to expect that the finite density environment brings about non-zero
$\gv$ even though we choose $\gv$ to be zero in the vacuum.

  There is no constraint at all for the choice of induced $\gv$ at
finite density.  We have no knowledge on even its sign.  Since we
regard $\gv$ in the present study as induced in dense quark matter,
the choice of $\gv$ has nothing to do with the vector meson property
in the vacuum.  [It might be related to an in-medium modification.]
It is presumably appropriate to measure the strength of $\gv$ in unit
of $\gs$, and we just try various values of $\gv$ to grasp a feeling
of its effect.

\begin{figure}
 \includegraphics[width=7.5cm]{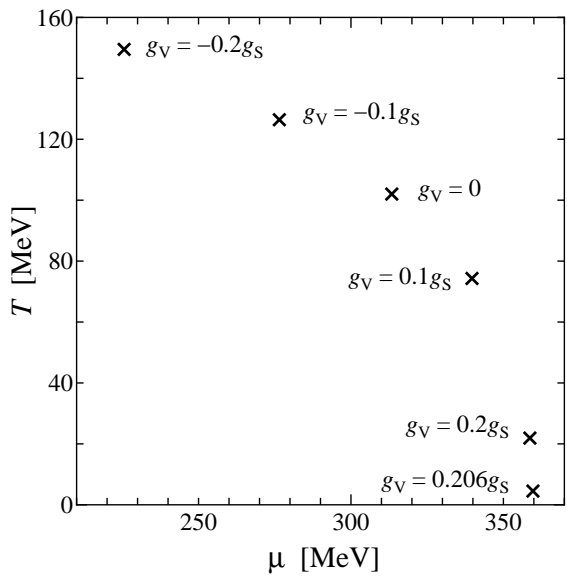}
 \caption{Dependence of the critical point location on the strength of
 the vector-channel interaction.}
 \label{fig:vector}
\end{figure}

  There are two modifications necessary to accommodate the
vector-channel interaction.  The condensation energy should be
$\Omega_{\text{cond}}\to\Omega_{\text{cond}}-\gv n_q^2$ where we
already defined $n_q$ in Eq.~(\ref{eq:nq}).  At the same time, the
quark chemical potential should be replace by the renormalized one,
\begin{equation}
 \mu_r = \mu - 2\gv n_q \,,
\end{equation}
like the quark mass replaced by the constituent one.  Then, we have to
solve the number constraint equation,
$n_q=-\partial\Op/\partial\mu_r$, together with the four gap equations
self-consistently.  In view of the condensation energy part, positive
$\gv$ seems to decrease the free energy for non-zero $n_q$, but the
chemical potential renormalization overcomes it and the free energy
becomes greater.  Because chiral symmetric phase has smaller $M_{ud}$
and thus larger $n_q$, the vector-channel interaction with $\gv>0$
delays chiral restoration.

  The results are summarized in Fig.~\ref{fig:vector} in the same way
as in Fig.~\ref{fig:nocep}.  It is remarkable that the qualitative
feature is quite similar to Fig.~\ref{fig:nocep}.  Thus, we can draw
the same conclusions as in the case of the $\UA$ anomaly term.

  The QCD critical point could be absent again in the case when $\gv$
is greater than around $0.206\gs$.  The critical value turns out to be
consistent with the known value in the two-flavor
case~\cite{Kitazawa:2002bc,Sasaki:2006ws}.  It should be noted that
the normalization of $\gs$ in
Refs.~\cite{Kitazawa:2002bc,Sasaki:2006ws} is different from the
present convention by a factor $2$.

  This value of critical $\gv$ is small as compared to the empirical
value suggested by the Fierz transformation.  If we take care of the
effect of the effective $\UA$ restoration, as we illustrate in
Fig.~\ref{fig:nocep}, the critical $\gv$ could be even smaller.


\section{CONCLUSIONS}

  We have formulated the $2+1$ flavor PNJL model with a simple ansatz
for the Polyakov loop effective potential.  We first confirmed that
our model setup works pretty well to account for recent results in the
zero-density lattice QCD simulation.  We then explored our perspective
toward the finite-density QCD phase transition.

  The phase diagram in our model study turns out to have three
(crossover) boundaries corresponding to $ud$-quark chiral restoration,
$s$-quark chiral restoration, and deconfinement characterized by the
Polyakov loop expectation value.  We have also computed the quark
number density and found that its behavior is governed by the
$u$-quark chiral condensate.  Our phase diagram is consistent with the
large $N_c$ argument and, in particular, we identified the phase
region with vanishing Polyakov loop and nonzero quark number density
as the quarkyonic phase.

  It would be intriguing to include the diquark condensates to
describe a family of the color superconducting phases.  The large
$N_c$ argument cannot access physics of color superconductivity, and
thus, nothing so far could predict the fate of the quarkyonic phase
region under the effect of color superconductivity.  One possibility
is that the quark-hadron continuity realizes at low temperature and
high density, and there appears crossover from the quarkyonic phase to
the color superconducting phase, which is to be interpreted as
crossover from confined to deconfined quark matter.  We remark that
this continuity scenario requires three flavors and there have already
been other scenarios (i.e.\ phase transitions with the coexisting
regions) within the PNJL model
framework~\cite{Roessner:2006xn,Abuki:2008ht}.

  Also, we have closely investigated parameter dependence of the
location of the QCD critical point.  We demonstrated that the QCD
critical point moves quite easily in accord to the choice of the $\UA$
anomaly strength and the vector-channel interaction.  Both are not
under theoretical control at finite density.  In fact, we have found
that the critical values of these parameters are within a conceivable
range in dense quark matter.  That means, not only the location but
also the existence of the QCD critical point is not robust at all in
the model study.

  Although we limited our discussions only to numerical results in
this paper, it could be viable to examine the density dependence of
the Columbia diagram in an analytical way~\cite{stephanov}.
Analytical understanding should be useful for the lattice QCD study
from the zero density approaching toward the critical point.

  To establish the existence or non-existence of the QCD critical
point, anyway, we must wait for future development of the
finite-density lattice simulation, or experimental confirmation.  [See
also Refs.~\cite{Hatta:2003wn,Asakawa:2008ti} for proposed
experimental signatures.]

  Finally, let us refer to some of recent attempts in the PNJL model.
The neutrality condition has been considered in
Ref.~\cite{Abuki:2008tx}.  It is known that the neutrality plays an
important role especially in bulk superconducting quark matter.
Because the (untraced) Polyakov loop is a color matrix, there arise
non-trivial coupling between the Polyakov loop and the color chemical
potential, which may bring a subtle difficulty involving the gauge
choice.  This is a future problem.  In
Ref.~\cite{Sakai:2008py,Sakai:2008um} the imaginary chemical potential
has been considered.  This is also an interesting future direction
toward the nature of high-density quark matter.

\acknowledgments

The author thanks Ph.~de~Forcrand, T.~Kunihiro, L.~McLerran,
R.~Redlich, C.~Sasaki, M.~A.~Stephanov, and W.~Weise for discussions.
This work is in part supported by Yukawa International Program for
Quark-Hadron Sciences.

\end{document}